\documentclass[11pt,twoside]{atmp}

\usepackage{amsmath,amssymb}

\usepackage[all]{xy}
\usepackage{color}
\usepackage{amsfonts}
\usepackage{latexsym}
\usepackage{slashed}

\newcommand{\be}{\begin{equation}}
\newcommand{\ee}{\end{equation}}
\newcommand{\bea}{\begin{eqnarray}}
\newcommand{\eea}{\end{eqnarray}}
\newcommand{\beg}{\begin{gather}}
\newcommand{\eeg}{\end{gather}}
\newcommand{\bb}{\mathbb}
\newcommand{\Tr}{\textrm{Tr}}
\newcommand{\tr}{\textrm{tr}}

\newcommand{\ph}{\phantom}
\newcommand{\mc}{\mathcal}
\newcommand{\bp}{\bar{\partial}}
\newcommand{\p}{\partial}

\begin{document}

\title[Spectrum of Non-K\"ahler Compactifications]
{Towards the Massless Spectrum of Non-K\"ahler Heterotic Compactifications}

\arxurl{hep-th/0605131}

\author[M. Cyrier and J. M. Lapan]{Michelle Cyrier and Joshua M. Lapan}

\address{Jefferson Physical Laboratory, Harvard University \\  Cambridge, MA  02138}  
\addressemail{cyrier@fas.harvard.edu, lapan@fas.harvard.edu}

\begin{abstract}

Fu and Yau constructed the first smooth family of gauge bundles over a class of non-K\"ahler, complex $3$-folds that are solutions to Strominger's system, the heterotic supersymmetry constraints with nonzero $H$-flux.  In this paper, we begin the study of the massless spectrum arising from compactification using this construction by counting zero modes of the linearized equations of motion for the gaugino in the supergravity approximation.  We rephrase the question in terms of a cohomology problem and show that for a trivial gauge bundle, this cohomology reduces to the Dolbeault cohomology of the $3$-fold, which we then compute.

\end{abstract}

\maketitle

\section{Introduction}

Heterotic string theory has long been known to have great promise for reproducing the standard model; unfortunately, amidst the excitement of branes, string dualities, and flux compactifications of type II and M-theory, it has been partially forgotten.  Compactifications of heterotic string theory that preserve $\mc{N}=1$ supersymmetry in four dimensions and with vanishing background flux were first studied in \cite{Candelas:1985en}; this was expanded to an analysis including nonzero $H$-flux by Strominger in \cite{Strominger:1986uh}.

In recent years, studies of compactifications of type II theories with various nonzero flux backgrounds have become common.  One of the most compelling reasons to study flux compactifications is that flux-free compactifications on Calabi-Yau manifolds lead to large moduli spaces corresponding to unconstrained scalar fields in the low-energy, four-dimensional description.  This is phenomenologically unsatisfying since it leaves us with a continuously infinite number of vacua.  On the other hand, it is well known that flux compactifications typically lift this degeneracy.  In fact, fluxes could conceivably be used to break supersymmetry or lift the cosmological constant to some positive value \`a la KKLT \cite{Kachru:2003aw}.

How flux compactifications of heterotic string theory achieve such noble goals is not yet well-understood from the spacetime/geometric perspective.  The difference between heterotic and type II theories is twofold: for one, we have the freedom to choose a gauge bundle which need not be the tangent bundle; second, there are no R-R fluxes to turn on, just the NS-NS $3$-form flux $H$.  The main difficulty in including $H$-flux is that it is encoded in the geometry of the internal manifold as torsion,\footnote{See \cite{LopesCardoso:2002hd} for a nice discussion on torsion.} meaning we have to deal with non-K\"ahler, though still complex, compactifications \cite{Strominger:1986uh}, \cite{Hull:1986kz}.

Giving up the K\"ahler condition destroys many nice results on K\"ahler geometry that we are accustomed to using.  For example, it is no longer generically true that de Rham cohomology is related to Dolbeault cohomology
\be
H^m_{dR}(K;\bb{R})^{\bb{C}} \ncong \bigoplus_{p+q = m} H^{p,q}_{\bp}(K;\bb{C}) .
\ee
We also have that
\be
H^{p,q}_{\bp}(K;\bb{C}) \ncong H^{q,p}_{\bp}(K;\bb{C}) .
\ee

\cutpage 
\setcounter{page}{2}

\noindent Another loss is that the Levi-Civita connection no longer annihilates the complex structure; instead, any connection annihilating the complex structure must contain torsion.  The Levi-Civita connection is the one commonly found in supergravity actions, and so one must be more careful when working with a non-K\"ahler compactification (see appendix \ref{non-kahler}, for example).  Other important properties of K\"ahler manifolds include the Lefschetz decomposition and the Hodge-Riemann bilinear relations \cite{Griffiths:1978}.

There are many other nice properties and theorems special to K\"ahler manifolds, and these play no small role in the dearth of studies of supersymmetric heterotic compactifications with $H$-flux.  Nevertheless, in recent years various groups have taken up this challenge for many reasons \cite{Dasgupta:1999ss,Becker:2002sx,Becker:2003dz,Becker:2003yv,Becker:2003gq,Becker:2003sh,Becker:2005nb,Gurrieri:2004dt,deCarlos:2005kh,Frey:2005zz,Manousselis:2005xa}, one of the most compelling reasons being the potential to lift moduli of the flux-free compactifications.

One might think that studying heterotic theory is moot since we expect any such study would be dual to some flux compactification of the better-understood type II or M-theory, but we should not forget that the use of duality is often that a difficult problem in one theory can be a simple one in the dual theory.  Even more compelling is that heterotic theory actually has a microscopic description in terms of $(0,2)$ conformal field theories, while in type II compactifications such a description does not yet exist.  If we fail to study flux compactifications of the heterotic theory we could be missing a wonderful opportunity, especially considering how natural heterotic theories seem when we are interested in reproducing properties of the standard model.

This paper considers properties of the massless spectrum of compactifications of heterotic supergravity using the construction recently developed by Fu and Yau \cite{Fu:2006vj}.  In their paper, Fu and Yau constructed gauge bundles over a sub-class of non-K\"ahler $3$-folds studied by Goldstein and Prokushkin \cite{Goldstein:2002pg} and proved the existence of solutions to Strominger's system.  We will refer to this construction as the FSY geometry and to the underlying $3$-fold as a GP manifold.  For a physical discussion and explicit examples of FSY geometries, see \cite{Becker:2006et}.

In section \ref{Strominger}, we review the constraints that Strominger derived on compactifications of heterotic supergravity preserving $\mc{N}=1$ supersymmetry in four dimensions \cite{Strominger:1986uh}.  In section \ref{GPFSY}, we review the constructions of Goldstein and Prokushkin and of Fu and Yau.  We then analyze the volume of the GP manifold in the FSY geometry as well as the volume of the fibers.  In section \ref{SUGRA}, we discuss the applicability of the supergravity approximation, compute the massless fields arising from the gaugino upon compactification, and find an explicit result for the choice of a trivial gauge bundle in terms of Hodge numbers of the GP manifold.  We compute the Hodge diamond of the GP manifold in section \ref{Hodge} and then discuss our results and future directions in section \ref{discussion}.

\section{Superstrings with Torsion}
\label{Strominger}

In \cite{Strominger:1986uh}, Strominger examined heterotic compactifications on warped product manifolds.  His compactification assumed a maximally symmetric four-dimensional spacetime $\mc{M}_4$ and internal six-dimensional manifold $K$ with metric
\be
\label{eqn:metric}
g^{0}_{MN}(x,y) = e^{-D(y)/2}\left(\begin{array}{cc}  g_{\mu\nu}(x)  &  0   \\   0  &  g_{mn}(y)  \end{array}\right),
\ee
where $x^\mu$ are coordinates on a patch of $\mc{M}_4$, $y^m$ are coordinates on a patch of $K$ (as we will soon see, $K$ must be complex and we will refer to the coordinates on a patch of $K$ as $\{z^a,\bar{z}^{\bar{a}}\}$), and capital Roman indices $M,N,\ldots$, are used for the full ten-dimensional spacetime.  To ensure a supersymmetric configuration, the supersymmetry variations of the fields must vanish.  This is trivially true for variations of the bosonic fields if we assume no fermionic condensates (see \cite{Manousselis:2005xa} for an example with condensates), so to preserve supersymmetry the variations of the fermionic fields must vanish.

After converting to string frame and applying some other simplifications, these variations yield the constraints\footnote{Throughout this note, we use the conventions: $H^{AS} = \frac{3}{2}H^{us}$, $\phi^{AS} = -\frac{1}{4}\phi^{us}$, where $AS$ refers to the conventions in \cite{Strominger:1986uh}.}
\begin{gather}
\nabla_M \epsilon - \frac{3}{8}H_M \epsilon = 0 ,       \nonumber \\
(\slashed{\nabla}\phi)\epsilon - \frac{1}{4}H\epsilon = 0 ,     \nonumber \\
\label{eqn:susy}
F_{MN} \Gamma^{MN} \epsilon = 0 ,
\end{gather}
where $H \equiv H_{MNP}\Gamma^{MNP}$ and $H_M \equiv H_{MNP}\Gamma^{NP}$.  Additionally, as required by anomaly cancellation, there is a modified Bianchi identity for the 3-form field strength $H$
\be
\label{eqn:bianchi}
\frac{3}{2}dH = \frac{\alpha'}{2}\left( \tr R\wedge R - \frac{1}{30}\Tr F \wedge F \right) ,
\ee
where $\tr$ is a trace over the vector representation of $O(1,9)$ and $\Tr$ is a trace over the adjoint represention of either $SO(32)$ or $E_8\times E_8$ (if we choose $SO(32)$, we can write this as simply a trace over the vector representation without the factor of $\frac{1}{30}$ \cite{Candelas:1985en}).  Also, $R$ refers to the Ricci 2-form of the Hermitian connection.\footnote{This connection has nonzero connection coefficients $\Gamma^a_{\ph{a}bc} = g^{a\bar{a}}g_{\bar{a}b,c}$ and $\Gamma^{\bar{a}}_{\ph{a}\bar{b}\bar{c}} = g^{\bar{a}a}g_{a\bar{b},\bar{c}}$.}  Finally, the warp factor is forced to be equal to the dilaton $D(y) = \phi(y)$.

The spinor $\epsilon$ is a ten-dimensional, Majorana-Weyl spinor, so we can decompose it as the sum of tensor products of four and six dimensional Weyl spinors, say $\epsilon = \epsilon_4 \otimes \eta + c.c.$.  Furthermore, the assumption of maximal symmetry in $\mc{M}_4$ requires that $F$ and $H$ have no components tangent to $\mc{M}_4$ and that they, as well as the dilaton $\phi$, only depend on the internal manifold $K$.  These facts simplify the constraints to
\be
\label{eqn:susy2}
\begin{array}{cc}
\nabla_\mu \epsilon_4 = 0 , \quad      &    \nabla_m \eta - \frac{3}{8}H_m \eta = 0 ,  \\
\\
(\nabla_m\phi)\gamma^m \eta -  \frac{1}{4} H\eta = 0 , \quad    &      F_{mn}\gamma^{mn}\eta = 0.
\end{array}
\ee
Thus, $\eta$ is covariantly constant with respect to a metric-compatible connection with torsion $\frac{3}{2}H$, the Strominger connection.\footnote{This is sometimes called the ``$H$-connection'' or the ``plus'' connection.}  There is an ambiguity in the Bianchi identity (\ref{eqn:bianchi}) in the choice of connection appearing in $R$.  In \cite{Fu:2006vj} and \cite{Becker:2006et}, the Hermitian connection is used and hence it is the one we mention below (\ref{eqn:bianchi}).  However, the connection with torsion $-\frac{3}{2}H$, referred to as the ``minus'' connection, is sometimes used \cite{Hull:1986kz}, \cite{Becker:2005nb}.  The ambiguity arises from a field redefinition in the effective action picture or from a choice of regularization scheme in the sigma model picture \cite{Sen:1986mg}.

Strominger showed in \cite{Strominger:1986uh} that $\eta$ could be used to construct an almost complex structure for $K$ ($J_m^{\ph{m}n} \equiv i \eta^\dag \gamma_m^{\ph{m}n}\eta$) that is $H$-covariantly constant
\be
\nabla_{m}J_{n}^{\ph{n}p} + \frac{3}{2}H_{\ph{p}ms}^p J_n^{\ph{n}s} - \frac{3}{2}H^s_{\ph{s}mn} J_s^{\ph{s}p} = 0
\ee
and has vanishing Nijenhuis tensor.  Thus, $J$ is integrable and is a complex structure for $K$; in fact, the metric $g_{mn}$ (\ref{eqn:metric}) is Hermitian with respect to $J$.  The supersymmetry constraints (\ref{eqn:susy}) also imply the existence of a nowhere-vanishing, holomorphic $(3,0)$-form $\Omega_{abc} = e^{-2\phi}\eta^\dag \gamma_{abc}\eta^*$, thus implying the vanishing of the first Chern class.  These conditions are equivalent to the existence of an $SU(3)$-structure.

In sum, Strominger recast (\ref{eqn:susy}) into the geometrical form:

\begin{enumerate}

\item $(K,g)$ must be a complex Hermitian manifold with vanishing first Chern class.

\item Using $J$ to denote the fundamental form in addition to the complex structure, we have
\be
\label{eqn:H-J}
\frac{3}{2}H = \frac{i}{2}(\bar{\partial} - \partial) J ;
\ee

\item $d^\dag J = i(\bp - \p)\ln ||\Omega||$;\footnote{Note that there was a sign error in \cite{Strominger:1986uh} that was corrected in \cite{Strominger:1990et}.}

\item $F$ is a $(1,1)$-form and must satisfy $J^{a\bar{b}}F_{a\bar{b}} = 0$;

\item finally, the modified Bianchi identity (\ref{eqn:bianchi}) must be satisfied.\footnote{For the rest of the paper, we will work in units where $\alpha'=1$.}

\end{enumerate}
These we refer to as ``Strominger's system'', which is the system of constraints that the FSY geometry was designed to solve.


\section{The GP Manifold and FSY Geometry}
\label{GPFSY}

\subsection{A Review}

In their paper \cite{Goldstein:2002pg}, Goldstein and Prokushkin gave an explicit construction of all complex $(n+1)$-folds that can be realized as principal holomorphic $T^2$ bundles over a complex $n$-fold.  In particular, they showed that if the base $2$-fold was Calabi-Yau, one could use this to construct a complex Hermitian $3$-fold satisfying constraints 1-3 of Strominger's system.  The remaining task for a heterotic solution was to construct a gauge bundle satisfying constraints 4-5.

Unlike the Calabi-Yau case, in non-K\"ahler compactifications one cannot embed the Strominger-connection in the gauge connection because the curvature form will have $(0,2)$ and $(2,0)$ components, so the choice of gauge bundle satisfying Strominger's system becomes much more complicated.  Fu and Yau undertook the difficult task of constructing just such a gauge bundle and were able to prove the existence of solutions to Strominger's system \cite{Fu:2006vj}.\footnote{In their original paper \cite{Fu:2005sm}, Fu and Yau proved the existence of a solution to the system of equations considered in section \ref{Strominger} but with opposite sign for (\ref{eqn:bianchi}).  In \cite{Fu:2006vj}, they have solved the system of equations from section \ref{Strominger}, which are the solutions relevant to heterotic compactifications.  The sign difference dates to a sign error in \cite{Strominger:1986uh}.  Fu and Yau have also considered a wider class of gauge bundles in the more recent paper.}  We briefly review these constructions here.

Let $M$ be a complex Hermitian 2-fold and choose
\be
\frac{\omega_P}{2\pi},\frac{\omega_Q}{2\pi}\in H^2(M;\bb{Z}) \cap \Lambda^{1,1}T^*M
\ee
(actually, Goldstein and Prokushkin only required that $\omega_P + i\omega_Q$ have no $(0,2)$-component, but Fu and Yau used the restriction that we have stated).  Being elements of integer cohomology, there are two unit-circle bundles over $M$, say $S^1_P$ and $S^1_Q$, whose curvature 2-forms are $\omega_P$ and $\omega_Q$, respectively.  Together, these form a $T^2$ bundle over $M$ which we will refer to as $K$, $K\stackrel{\pi}{\rightarrow} M$.

Given this setup, Goldstein and Prokushkin showed that if $M$ admits a non-vanishing, holomorphic $(2,0)$-form, then $K$ admits a non-vanishing, holomorphic $(3,0)$-form.  Furthermore, they showed that if $\omega_P$ or $\omega_Q$ are nontrivial in cohomology on $M$, then $K$ admits \emph{no} K\"ahler metric.  They were able to construct the non-vanishing holomorphic $(3,0)$-form and a Hermitian metric on $K$ simply from data on $M$.  In particular, for the choice $M=K3$, they were able to compute the Betti numbers of $K$, as well as $h^{0,1}$ and $h^{1,0}$.

The curvature 2-form $\omega_P$ determines a non-unique connection $\nabla$ on $S^1_P$ (and similarly for $\omega_Q$ and $S^1_Q$).  A connection determines a split of $TK$ into a vertical and horizontal subbundle---the horizontal subbundle is composed of the elements of $TK$ that are annihilated by the connection $1$-form, the vertical subbundle is then, roughly speaking, the elements of $TK$ tangent to the fibers.  Over an open subset $U\subset M$, we have a local trivialization of $K$ and we can use unit-norm sections, $\xi$ of $S^1_P$ and $\zeta$ of $S^1_Q$, to define local coordinates for $z\in U\times T^2$ by
\be
\label{eqn:coords}
z = (p,e^{ix}\xi(p),e^{iy}\zeta(p)) , 
\ee
where $p = \pi(z)\in U$.  The sections $\xi$ and $\zeta$ also define connection 1-forms via
\be
\nabla \xi = i\alpha\otimes \xi      \qquad     \textrm{and}     \qquad     \nabla \zeta = i\beta\otimes\zeta,
\ee
where $\omega_P = d\alpha$ and $\omega_Q = d\beta$ on $U$, and $\alpha$ and $\beta$ are necessarily real to preserve the unit-norms of $\xi$ and $\zeta$.

The complex structure is given on the fibers by $\partial_x\rightarrow\partial_y$ and $\partial_y\rightarrow -\partial_x$ while on the horizontal distribution it is induced by projection onto $M$ (actually, this just gives an almost complex structure, but Goldstein and Prokushkin proved that it is integrable \cite{Goldstein:2002pg}).  Given a Hermitian $2$-form $\omega_M$ on $M$, the 2-form
\be
\label{eqn:hermitian form}
\omega_u = \pi^*\left(e^{u} \omega_M\right) + (dx+\pi^*\alpha)\wedge(dy+\pi^*\beta),
\ee
where $u$ is some smooth function on $M$, is a Hermitian $2$-form on $K$ with respect to this complex structure.  The connection $1$-form
\be
\rho = (dx+\pi^*\alpha) + i(dy+\pi^*\beta)
\ee
annihilates elements of the horizontal distribution of $TK$ while reducing to $dx+idy$ along the fibers.  These data define the complex Hermitian $3$-fold $(K,\omega_u)$, which we call the GP manifold \cite{Goldstein:2002pg}.

Fu and Yau undertook the more difficult problem of proving the existence of gauge bundles over the GP manifold with Hermitian-Yang-Mills connections satisfying the Bianchi identity (\ref{eqn:bianchi}).  They took the Hermitian form (\ref{eqn:hermitian form}) and converted the Bianchi identity into a differential equation for the function $u$.  Under the assumption
\be
\label{eqn:u assumptions}
\left(\int_{K3} e^{-4u}\frac{\omega_{K3}^2}{2} \right)^{1/4} \ll 1 = \int_{K3} \frac{\omega_{K3}^2}{2} ,
\ee
they then specialized to a $K3$ base and showed that there exists a solution $u$ to the Bianchi identity for \emph{any} compatible choice of gauge bundle $V$ and curvatures $\omega_P$ and $\omega_Q$\footnote{See equation (\ref{eqn:bundle constraint}) for an explanation.} such that the gauge bundle $V$ over $K$ is the pullback of a stable, degree 0 bundle $E$ over $K3$, $V = \pi^*E$ \cite{Fu:2006vj}; this is what we call the FSY geometry.  In \cite{Fu:2006vj} and \cite{Becker:2006et}, it was shown that no such solution exists for a $T^4$ base.  This is in agreement with arguments from string duality ruling out the Iwasawa manifold as a solution to the heterotic supersymmetry constraints \cite{Gauntlett:2003cy}.

\subsection{The Volume}

\subsubsection{Of the Total Space}

From here on, we will take the base $M$ to be $K3$.  In this paper, we will be concerning ourselves with questions relating to supergravity compactifications on the FSY geometry.  We must therefore understand the curvature scales involved in the construction.  One issue that we can address is the overall volume of $K$
\be
\textrm{Vol}(K) \propto \int_{K} \omega_u^3 .
\ee
Let $\{U_\alpha\}$ be a good open cover of $M$ with subordinate partition of unity $\{\rho_\alpha\}$.  Then $\{V_\alpha := \pi^{-1}(U_\alpha)\}$ is a good open cover of $K$ with induced subordinate partition of unity $\tilde{\rho}_\alpha$ that is constant along the fibers, so $\tilde{\rho}_\alpha\pi^* = \pi^*\rho_\alpha$.  Furthermore, we have the local trivialization $V_\alpha \stackrel{\phi_\alpha^{-1}}{\cong} U_\alpha\times T^2$, so we have
\be
\int_{K}\omega_u^3 = \sum_\alpha \int_{U_\alpha\times T^2} \phi^*_\alpha (\tilde{\rho}_\alpha \omega_u^3)  .
\ee

In fact, we may choose the $\phi_\alpha$ so that the vertical subbundle of $TV_\alpha \cong TU_{\alpha}\times TT^2$ is mapped isomorphically to $U_\alpha\times TT^2$ and similarly for the horizontal subbundle of $TV_\alpha$ to $TU_\alpha\times T^2$.  We also have the inclusion $\iota_\alpha:~ T^2 \hookrightarrow U_\alpha \times T^2$ which induces the trivial projection $\iota_\alpha^*:~ \Omega^*(U_\alpha\times T^2) \rightarrow \Omega^*(T^2)$.  Now, $\omega_u^3$ is a sum of terms of the form $\pi^*\gamma \wedge \lambda$, where $\gamma\in\Omega^*(U_\alpha)$ and $\lambda$ annihilates elements of the horizontal subbundle of $TV_\alpha$.  Then we find
\be
\int_{U_\alpha\times T^2} \phi_\alpha^*(\pi^*\gamma \wedge \lambda)  =  \left( \int_{U_\alpha} \gamma \right) \left( \int_{T^2} \iota^*(\phi_\alpha^*(\lambda)) \right) .
\ee
We find for the volume of $K$
\be
\int_{K}\omega_u^3 = \sum_\alpha \left( \int_{U_\alpha} \rho_\alpha e^{2u}\omega_M^2 \right) \left( \int_{T^2} dx\wedge dy \right) \propto \int_M e^{2u}\omega_M^2 \gg 1 ,
\ee
which follows from (\ref{eqn:u assumptions}), so $K$ is large enough for the supergravity approximation to be valid.  However, we should also check the volume of the $T^2$ fibers.

\subsubsection{Of the Fibers}

In the GP manifold, the $T^2$ fibers are taken to have size $4\pi^2$.  This is simply because the coordinates $x$ and $y$ were defined such that they have periodicity $2\pi$ (\ref{eqn:coords}).  If we rescale both by an integer $N$, the form defining the horizontal distribution becomes
\be
\rho_N = \left(dx + \frac{\pi^*\alpha}{N}\right) + i\left(dy + \frac{\pi^*\beta}{N}\right) .
\ee
This normalization of $\rho_N$ preserves the property that $\frac{i}{2}\rho_N\wedge\bar{\rho}_N$ restricted to the fibers is just $dx\wedge dy$, so the Hermitian form (\ref{eqn:hermitian form}) keeps the same form with $\rho$ replaced by $\rho_N$.

In the Bianchi identity (\ref{eqn:bianchi}), the only place in which $\omega_P$ and $\omega_Q$ enter is through the Hermitian form and so rescaling the $T^2$ is, as far as the Bianchi identity is concerned, equivalent to keeping the volume of the $T^2$ fixed and instead rescaling $\omega_P$ and $\omega_Q$ each by $N$.  More generally, we find that these two setups produce the same solution for the function $u$:
\be
\begin{array}{ccc}
\begin{array}{c}  \textrm{Vol}(T^2) = 4\pi^2  \\  \textrm{Curvatures: } N\omega_P, M\omega_Q  \end{array}     &
\longleftrightarrow       &
\begin{array}{c}  \textrm{Vol}(T^2) = 4\pi^2NM  \\  \textrm{Curvatures: } \omega_P, \omega_Q  \end{array}  
\end{array}
\ee
where $N,M\in\bb{Z}^+$.

It would seem, then, that we are free to rescale the $T^2$ to be arbitrarily large.  However, there is a constraint pointed out in \cite{Fu:2006vj} and \cite{Becker:2006et} which restricts $\omega_P$ and $\omega_Q$ quite heavily.  The constraint comes from integrating the Bianchi identity and is
\be
\label{eqn:bundle constraint}
24 - Ch_2(E)  = \int_{K3} \left( \left|\left|\frac{\omega_P}{2\pi}\right|\right|^2 + \left|\left|\frac{\omega_Q}{2\pi}\right|\right|^2 \right)\frac{\omega_M^2}{2} ,
\ee
where $Ch_2(E)$ is the integral of the second Chern character ($\tr F^2$) over $K3$, and similarly $24$ comes from integrating $\tr R_{K3}^2$ which yields the Euler characteristic of $K3$.

Furthermore, $\int_{K3} \left|\left|\frac{\omega_P}{2\pi}\right|\right|^2 \frac{\omega_M^2}{2}$ can be computed using the intersection form on $K3$ and is known to be a positive, even integer.  It therefore seems we cannot make the volume of the $T^2$ significantly larger than the string scale.  However, this statement is not quite right; as Goldstein and Prokushkin note \cite{Goldstein:2002pg}, even if $\omega_P$ or $\omega_Q$ (but not both) is trivial in cohomology, the $3$-fold is still non-K\"ahler.  There is nothing to prevent us from considering models in which one of the circle bundles, say $S^1_P$, is trivial.  In these cases, since $\int_{K3} \left|\left|\frac{\omega_P}{2\pi}\right|\right|^2 \frac{\omega_M^2}{2} = 0$, we are free to rescale that circle to our heart's content.  We are then left with just one circle that cannot be made much larger than the string scale.  The correct statement, then, is that we cannot make both circles arbitrarily large.


\section{Heterotic Supergravity}
\label{SUGRA}

We see that the volume of the $K3$ is large but the volume of the $T^2$ fibers is generically of order the string scale.  This is a problem for a simple KK reduction of the ten-dimensional supergravity, but not because of curvature scales; rather, it is because we know that nonzero winding and momentum modes of the string become light when the $T^2$ is of order $\alpha'$.  This can be simply remedied by including these new light degrees of freedom in the dimensionally-reduced action.  We will leave this for future work and just work with the compactification of the ten-dimensional effective action below.  Note also that large curvatures associated with the gauge bundle can become problematic for a supergravity approximation, but we see from (\ref{eqn:bundle constraint}) that the curvature is bounded above since the right-hand side of the equation is non-negative.

\subsection{Linearized EOM's}

The string-frame action is (see appendix \ref{string-frame}):
\bea
L^{(S)}_{Het} & = & -\textstyle \frac{1}{2} e^{-2\phi}\sqrt{-G} \Big[  \frac{1}{\kappa^2} R 
- \frac{4}{\kappa^2}D_M\phi D^M\phi 
+ \frac{1}{2\kappa^2} \Tr(F^2) 
+ \frac{3}{4\kappa^2} H^2      \nonumber  \\
& + & \textstyle \bar{\psi}_M \Gamma^{MNP} D_N \psi_P 
+ \bar{\psi}_M \Gamma^{MP}\Gamma^N \psi_P D_N\phi 
+ \bar{\lambda}\Gamma^M D_M \lambda     \nonumber \\
& - & \bar{\lambda}\Gamma^M\lambda D_M\phi    
+ \textstyle \Tr\big( \bar{\chi}\Gamma^M D_M\chi - \bar{\chi}\Gamma^M \chi D_M\phi \big)       \nonumber \\ 
& + & \frac{1}{2}\Tr\big( \bar{\chi}\Gamma^M\Gamma^{NP} (\psi_M + \frac{\sqrt{2}}{12}\Gamma_M\lambda) F_{NP} \big)     
+ \textstyle \frac{1}{\sqrt{2}}\bar{\psi}_M\Gamma^N\Gamma^M\lambda D_N\phi     \nonumber \\
& - & \frac{1}{8}\Tr\big( \bar{\chi}\Gamma^{MNP}\chi \big) H_{MNP} 
- \frac{1}{8}\Big( \bar{\psi}_M\Gamma^{MNPQR}\psi_R     
+ \textstyle 6\bar{\psi}^N\Gamma^P\psi^Q      \nonumber \\
& - & \sqrt{2}\bar{\psi}_M\Gamma^{NPQ}\Gamma^M\lambda \Big) H_{NPQ}  +   (\textrm{Fermions})^4 .
\eea
Decomposing $\chi = \chi^I T^I$ and $F_{MN} = F^I_{MN} T^I$, where $T^I$ are generators of the gauge group satisfying $\Tr(T^I T^J) = \delta^{IJ}$, we find the linearized equations of motion for the fermions:

\textbf{Gaugino:}
\bea
0 & = & 2\Gamma^M D_M \chi^I - 2 \Gamma^M\chi^I D_M\phi + \frac{1}{2}\Gamma^M \Gamma^{NP} \psi_M F^I_{NP}  + \frac{\sqrt{2}}{3}\Gamma^{NP}\lambda F^I_{NP}     \nonumber \\
& & - \frac{1}{4}\Gamma^{MNP}\chi^I H_{MNP}
\eea

\textbf{Dilatino:}
\bea
0 & = & 2\Gamma^M D_M\lambda - 2\Gamma^M \lambda D_M\phi - \frac{\sqrt{2}}{3}\Gamma^{NP}\chi^I F^I_{NP} + \frac{1}{\sqrt{2}}\Gamma^M\Gamma^N\psi_M D_N\phi      \nonumber \\
& & - \frac{\sqrt{2}}{8}\Gamma^M\Gamma^{NPQ}\psi_M H_{NPQ}
\eea

\textbf{Gravitino:}
\bea
0 & = & 2\Gamma^{MNP}D_N\psi_P + 2\Gamma^{MP}\Gamma^N\psi_P D_N\phi + \frac{1}{2} \Gamma^{NP}\Gamma^M\chi^I F^I_{NP}      \nonumber \\
& & + \frac{1}{\sqrt{2}}\Gamma^N\Gamma^M\lambda D_N\phi - \frac{1}{4}\Gamma^{MNPQR}\psi_R H_{NPQ} - \frac{3}{2}\Gamma_N\psi_P H^{MNP}     \nonumber \\
& & + \frac{\sqrt{2}}{8}\Gamma^{NPQ}\Gamma^M\lambda H_{NPQ}.
\eea

Strominger's solution \cite{Strominger:1986uh}, and also the solution in the preceding paper \cite{Candelas:1985en}, trivially satisfied these equations of motion by assuming no fermionic condensates.  We are interested in the four-dimensional effective theory, in particular the massless spectrum, arising from compactifications on the FSY geometry.  Since we know we will have supersymmetry in the four-dimensional theory, we can just look for variations of the fermionic fields satisfying the equations of motion while holding the bosonic fields fixed; the massless bosonic fields will then simply be superpartners of the massless fermionic fields.

There is, of course, a limitation in our methodology.  We have ignored higher order $\alpha'$ corrections and a superpotential that is expected to be generated (see \cite{Becker:2003dz}, for example)---these should lift some of the massless fields.  We expect that this method will ultimately provide an upper bound to the number of massless fields, so let us proceed with it.

In ten dimensions, $\mc{N}=1$ supersymmetry implies that we have one Majorana-Weyl spinor supercharge.  We can decompose a ten-dimensional Majorana-Weyl spinor as
\be
\epsilon^{(10)}_- = \epsilon^{(4)}_- \otimes \epsilon^{(6)}_+ + \epsilon^{(4)}_+ \otimes \epsilon^{(6)}_-
\ee
where $\epsilon^{(4)}_\pm$ are four-dimensional, charge conjugate Weyl spinors, while $\epsilon^{(6)}_\pm$ are six-dimensional, charge conjugate Weyl spinors.  We take as our convection for the ten-dimensional gamma matrices:
\be
\Gamma^\mu = \gamma^\mu \otimes \mathbf{1}    \qquad    \textrm{and}    \qquad    \Gamma^m = \gamma_5 \otimes \gamma^m,
\ee
where $\gamma^\mu$ are the four-dimensional gamma matrices, $\gamma^m$ the six-dimensional ones, and $\gamma_5$ is the four-dimensional chirality operator.  The six-dimensional, $H$-covariantly constant spinor $\eta$ (\ref{eqn:susy2}) and its charge conjugate $\eta^*$ then satisfy $\gamma^{\bar{a}}\eta = 0 = \gamma^a \eta^*$, which in fact implies that $\eta$ has positive chirality and $\eta^*$ has negative chirality.

Note that the set $\{\eta, \gamma^a\eta, \gamma^{ab}\eta, \gamma^{abc}\eta\}$ spans the space of six-dimensional spinors.  The last one, $\gamma^{abc}\eta$, is the only one annihilated by all the $\gamma^a$'s, so it should be proportional to $\eta^*$.  In particular,
\be
\label{eqn:eta*}
\eta^* = \frac{1}{\sqrt{48}}e^{2\phi}\Omega_{abc}\gamma^{abc}\eta
\ee
up to an overall phase.  It is also true that $\eta^*$ is $H$-covariantly constant, which follows by noting that $e^{2\phi}\Omega_{abc}$ is $H$-covariantly constant.


\subsection{Counting the Massless Gauginos}
\label{Gaugino}

We can use (\ref{eqn:eta*}) and the basis $\{\eta,\gamma^a\eta,\gamma^{ab}\eta,\gamma^{abc}\eta\}$ to write the most general Ansatz for the variation of the gaugino as
\be
\delta\chi = \epsilon_- \otimes \left( C \eta + C_{ab}\gamma^{ab}\eta \right) - \epsilon_+ \otimes \left( \bar{C} \eta^* + \bar{C}_{\bar{a}\bar{b}} \gamma^{\bar{a}\bar{b}} \eta^* \right) ,
\ee
where $C,C_{ab}\in\Omega^*(K;V)$ are forms valued in some representation $V$ of the gauge group.  This is the most general form since $\chi$ must be a Majorana-Weyl ten-dimensional spinor.  See appendix A of \cite{Manousselis:2005xa} for details on this.

When we choose a gauge bundle with structure group $G$ and embed it in $E_8\times E_8$ or $SO(32)$, the adjoint will decompose into a sum of products of representations of the smaller groups.  One of these terms will transform as an adjoint of $G$ and we will ignore variations of this term since it is the one that couples to the other fermions.  This simplifies our lives by allowing us to consider the variation of the other portions of the gaugino independent from the other fermions.  This implies the gaugino equation of motion takes the form
\be
\label{eqn:simplified-gaugino}
0 = 2 \left( D_a C + 4 D^b C_{ba} - 4 \partial^b \phi C_{ba} \right) \gamma^a \eta + 2(D_a C_{bc})\gamma^{abc}\eta,
\ee
where we recall that $D$ is the Levi-Civita connection plus the gauge connection---see appendix \ref{derivation} for the derivation of (\ref{eqn:simplified-gaugino}).

By rescaling the $C$'s by $e^{\phi/2}$, the equations take the form
\bea
\label{eqn:gaugino-eom}
0 & = & D_a C + \frac{1}{2}\partial_a\phi C + 4 D^b C_{ba} - 2\partial^b\phi C_{ba}       \nonumber \\
0 & = & D_{[a}C_{bc]} + \frac{1}{2}\partial_{[a}\phi C_{bc]} ,
\eea
or by writing $C_{(0)} = C$ and $C_{(2)} = \frac{1}{2}C_{ab}dz^a\wedge dz^b$ we can recast these equations as
\be
0 = \mc{D} C_{(0)} + 4\mc{D}^\dag C_{(2)}       \qquad     \textrm{and}     \qquad      0 = \mc{D}C_{(2)}.
\ee
This defines the differential operator $\mc{D}:~\Omega^{p,q}(K;V)\rightarrow\Omega^{p+1,q}(K;V)$, while the adjoint is defined via the inner product
\be
(\alpha,\beta) := \int_K \alpha^\dag\wedge\beta,
\ee
where $\alpha^\dag = \bar{\alpha}^T$ takes values in the dual vector bundle to $V$.

$\mc{D}^2$ is just the $(2,0)$ part the curvature 2-form of the gauge bundle, which is required to be a $(1,1)$-form by supersymmetry, so $\mc{D}^2 = 0$ and similarly $\mc{D}^{\dag 2}=0$.  We then find that
\be
0 = \Delta_{\mc{D}} C_{(0)}    \qquad        \textrm{and}        \qquad       0 = \Delta_{\mc{D}} C_{(2)}
\ee
so that the $C$'s are $\mc{D}$-harmonic forms.  The space of solutions to (\ref{eqn:gaugino-eom}) is then spanned by these forms, reducing the question of counting massless gaugino modes to the question of computing the dimensions of the $\mc{D}$-cohomology groups $H^{0,0}_{\mc{D}}(K;V)$ and $H^{2,0}_{\mc{D}}(K;V)$.  Furthermore, since the dilaton $\phi$ is continous over the compact manifold $K$, the cohomology of $e^{-a\phi}\mc{D}e^{a\phi}$ (for constant $a$) is the same as that of $\mc{D}$, meaning we can rescale the $C$'s to consider equations of the form $\mc{D}C + a(d\phi)\wedge C = 0$ for any $a$.  In particular, we can choose $a$ to eliminate the above $\phi$ dependence so that $\mathcal{D}$ will become a standard twisted Dolbeault operator.

These cohomologies are rather abstract and we cannot simplify things as in the Calabi-Yau case by embedding the Strominger connection in the gauge connection.  The reason for this is that the field strength must be a $(1,1)$-form by the supersymmetry constraints, but the Ricci $2$-form will not be purely $(1,1)$ for the Strominger connection.  However, we do have the simpler option of choosing the bundle to be trivial, as explained in \cite{Becker:2006et}.\footnote{There is also an example of a nontrivial gauge bundle presented in \cite{Becker:2006et}, along with a proof that all stable bundles over the GP manifold, $K$, satisfying Strominger's system must be a bundle pulled back from the $K3$ tensored with a line bundle over $K$.}  A trivial line bundle is semi-stable, not stable; however, this is not a problem since the main usage of stability appears in the Donaldson-Uhlenbeck-Yau theorem, which proves the existence of a connection that yields a $(1,1)$ curvature form satisfying $F_{a\bar{b}}g^{a\bar{b}} = 0$.  In the case of a trivial bundle, these conditions are obviously met and so semi-stable will suffice.

For the choice of trivial line bundle, the twisted cohomology problem reduces to that of the $\partial$ operator, or if we consider the complex conjugate equations, the solutions are given by the usual Dolbeault cohomology groups $H^{0,0}_{\bp}(K;\bb{C})$ and $H^{0,2}_{\bp}(K;\bb{C})$.  As we will see in the next section, $h^{0,0} = 1 = h^{0,2}$, so for this choice of gauge group we get two massless fermions transforming in the adjoint of $E_8\times E_8$ or $SO(32)$.

\subsection{A Quick Check}

As a check on this method, let us consider what happens in the Calabi-Yau case using standard embedding.  Under $E_8\times E_8 \rightarrow SU(3)\times E_6\times E_8$, the adjoint $\mathbf{(248,1)+(1,248)}$ decomposes as
\be
\mathbf{(1,78,1) + (1,1,248) + (8,1,1) + (3,27,1) + (\bar{3},\overline{27},1)} .
\ee
As mentioned earlier, we will focus on the variations of the gaugino other than the adjoint of $SU(3)$, $\mathbf{(8,1,1)}$.  First, the adjoint of $E_6\times E_8$, $\mathbf{(1,78,1)}+\mathbf{(1,1,248)}$, are scalars as far as the $SU(3)$ connection is concerned, so we get $h^{0,0}(CY_3)+h^{0,2}(CY_3) = 1$ fermion transforming in the adjoint of $E_6\times E_8$.

For the $\mathbf{(3,27,1)}$, the equations take the form
\be
\label{eqn:CY-eqn}
0 = D_a C^{d} + 4 D^b C_{ba}^{\ph{ba}d} ,     \qquad     0  =  D_{[a}C_{bc]}^{\ph{bc}d} ,
\ee
where we have suppressed indices for $E_6\times E_8$.  We can use the metric to lower the holomorphic index $d$ to an antiholomorphic index, thus leaving us with $h^{1,0}(CY_3) + h^{1,2}(CY_3) = h^{1,2}(CY_3)$ fermions transforming in the $\mathbf{27}$ of $E_6$.  Finally for the $\mathbf{(\bar{3},\overline{27},1)}$, we have an equation similar to (\ref{eqn:CY-eqn}) except with an antiholomorphic index $\bar{a}$ in place of $d$.  Lowering $\bar{a}$ to a holomorphic index using the metric is not useful since it will not be antisymmeterized with the other indices.  However, we can contract $\bar{a}$ with one index from the covariantly-constant, antiholomorphic $(0,3)$-form $\bar{\Omega}$, yielding $h^{2,0}(CY_3)+h^{2,2}(CY_3) = h^{1,1}(CY_3)$ fermions transforming in the $\mathbf{\overline{27}}$ of $E_6$.  These results are in agreement with the well-known counting of massless modes arising from the connection $1$-form, $A_M$.  See, for example, \cite{Polchinski:1998rr}.


\section{Computing the Hodge Diamond}
\label{Hodge}

Goldstein and Prokushkin \cite{Goldstein:2002pg} explained a method for computing the Hodge numbers and used it to compute $h^{0,1}$ and $h^{1,0}$.  They showed that the Dolbeault cohomology groups $H^{p,q}_{\bar{\partial}}(K)$ are left invariant by the actions of $\partial_x$ and $\partial_y$, which literally means that if we move an element of $H^{p,q}_{\bar{\partial}}(K)$ around a fiber $\pi^{-1}(p)$, $p\in K3$, the form should remain constant.  This means we can express all forms as a sum of wedge products of $\rho$, $\bar{\rho}$, and forms pulled back from the $K3$.

We should note, before proceeding, that we will restrict attention to the case where $\omega_P$ and $\omega_Q$ are anti-selfdual $(1,1)$-forms.  The construction in \cite{Goldstein:2002pg} only requires that $\omega_P + i\omega_Q$ have no $(0,2)$-component and that they each have anti-selfdual $(1,1)$-components.  Fu and Yau restrict to GP manifolds where $\omega_P$ and $\omega_Q$ are $(1,1)$-forms \cite{Fu:2006vj}, so the computations that follow hold for the case considered by Fu and Yau but do not encompass all manifolds constructed by Goldstein and Prokushkin.

Let us review the computation of $h^{1,0}$ from \cite{Goldstein:2002pg} for illustration first.  If $\xi\in H^{1,0}_{\bar{\partial}}(K)$, then
\be
\xi = (\pi^* s)\rho + \pi^* s^{1,0} ,
\ee
where $s$ is a function on $K3$ and $s^{1,0}$ is a $(1,0)$-form on $K3$.  Recall that $\bar{\partial}\rho = \pi^*(\omega_P + i\omega_Q)$ and $\bp\bar{\rho} = 0$.  Thus, $0 = \bp\xi = \pi^*(\bp s)\wedge\rho + \pi^*\left( s(\omega_P + i\omega_Q) + \bp s^{1,0} \right)$.  The first term tells us $\bp s = 0$, which means that $s$ must be a constant.  The term pulled back from $K3$ tells us that $s(\omega_P + i\omega_Q) = -\bp s^{1,0}$, but since $\omega_P + i\omega_Q$ is assumed to be nontrivial in the Dolbeault cohomology of $K3$, then the only solution to this is $s = 0$ and $\bp s^{1,0} = 0$.  Thus, we have the result $h^{1,0}(K) = h^{1,0}(K3) = 0$.  Similarly, Goldstein and Prokushkin found $h^{0,1}(K) = h^{0,1}(K3) + 1 = 1$.

Note that the Hodge theorem implies that even for non-K\"ahler manifolds, $H^{p,q}_{\bp}(K) = H^{n-p,n-q}_{\bp}(K)$ \cite{Griffiths:1978}, however $H^{p,q}_{\bp}(K) \neq H^{q,p}_{\bp}(K)$.  In any event, we only have to compute some of the Hodge numbers.  If $\xi\in H^{1,1}_{\bp}(K)$, then
\be
\xi = (\pi^*s)\rho\wedge\bar{\rho} + \rho\wedge\pi^*s^{0,1} + \bar{\rho}\wedge\pi^*s^{1,0} + \pi^*s^{1,1} .
\ee
Requiring $\bp\xi = 0$ implies
\begin{gather}
\bp s = 0 ,   \qquad   s(\omega_P + i\omega_Q) - \bp s^{1,0} = 0 ,     \qquad    \bp s^{0,1} = 0 ,        \nonumber \\
\textrm{and}   \quad    (\omega_P + i\omega_Q)\wedge s^{0,1} + \bp s^{1,1} = 0 .
\end{gather}
As above, we find: $s = 0$; $\bp s^{1,0} = 0$, which then implies $s^{1,0} = 0$ ($h^{1,0}(K3) = 0$); and $s^{0,1} = \bp t$ ($h^{0,1}(K3) = 0$), where $t$ is a function on $K3$, which then implies $s^{1,1} = t^{1,1} - t(\omega_P + i\omega_Q)$, where $\bp t^{1,1} = 0$.  So we have
\be
\xi = \rho\wedge\bp\pi^*t + \pi^* \left( t^{1,1} - t(\omega_P + i\omega_Q) \right) = \pi^* t^{1,1} - \bp \left( (\pi^*t)\rho  \right) .
\ee
This last term is exact, and $\pi^*(t^{1,1}+\bp u^{1,0}) = \pi^*t^{1,1} + \bp\pi^*u^{1,0}$, so we find $h^{1,1}(K) = h^{1,1}(K3) = 20$.

Now take $\xi\in H^{2,0}_{\bp}(K)$, so we have
\be
\xi = \rho\wedge\pi^*s^{1,0} + \pi^*s^{2,0} .
\ee
$\bp\xi = 0$ implies $\bp s^{1,0} = 0$, so $s^{1,0} = 0$, which in turn implies that $\bp s^{2,0} = 0$, so $s^{2,0} = c\Omega^{2,0}_{K3}$, where $c$ is a constant and $\Omega^{2,0}_{K3}$ is the nowhere-vanishing, holomorphic $(2,0)$-form on $K3$.  Thus, $h^{2,0}(K) = 1$.

If $\xi\in H^{0,2}_{\bp}(K)$, then
\be
\xi = \bar{\rho}\wedge\pi^*s^{0,1} + \pi^*s^{0,2} .
\ee
Then $\bp\xi = 0$ implies that $\bp s^{0,1} = \bp s^{0,2} = 0$.  Shifting $s^{0,1}$ or $s^{0,2}$ by a $\bp$-exact form just shifts $\xi$ by a $\bp$-exact form, so $h^{0,2}(K) = h^{0,1}(K3) + h^{0,2}(K3) = 1$.

Finally, suppose $\xi\in H^{1,2}_{\bp}(K)$, then
\be
\xi = \rho\wedge\bar{\rho}\wedge\pi^*s^{0,1} + \bar{\rho}\wedge\pi^*s^{1,1} + \rho\wedge\pi^*s^{0,2} + \pi^*s^{1,2} .
\ee
Requiring $\bp\xi = 0$ implies
\begin{gather}
\bp s^{0,1} = 0 ,     \qquad     (\omega_P + i\omega_Q)\wedge s^{0,1} - \bp s^{1,1} = 0 ,       \qquad       \bp s^{0,2} = 0 ,         \nonumber \\
\textrm{and}    \quad     (\omega_P + i\omega_Q)\wedge s^{0,2} + \bp s^{1,2} = 0 ;
\end{gather}
however, these last two equations are trivially true since $K3$ is a complex 2-fold.  These translate into: $s^{0,1} = \bp t$; $s^{1,1} = t^{1,1} + t(\omega_P + i\omega_Q)$, where $\bp t^{1,1} = 0$; $s^{1,2} = \bp u^{1,1}$ ($h^{1,2}(K3) = 0$); and $s^{0,2} = c\bar{\Omega}^{0,2}_{K3} + \bp t^{0,1}$, where $\bar{\Omega}^{0,2}_{K3}$ is the complex conjugate of the holomorphic $(2,0)$-form on $K3$ and $c$ is a constant.  So we have now
\bea
\xi & = & \rho\wedge\bar{\rho}\wedge\bp\pi^*t + \bar{\rho}\wedge\pi^*t^{1,1} + \pi^*(\omega_P + i\omega_Q)\wedge\bar{\rho} (\pi^*t)      \nonumber \\
& & + c\rho\wedge\pi^*\bar{\Omega}^{0,2}_{K3} + \rho\wedge\bp\pi^*t^{0,1} + \bp\pi^*u^{1,1}     \nonumber \\
& = & \bp\left( (\pi^*t)\rho\wedge\bar{\rho} \right) + \bar{\rho}\wedge\pi^*t^{1,1} + c\rho\wedge\pi^*\bar{\Omega}^{0,2}_{K3} - \bp \left( \rho\wedge\pi^*t^{0,1} \right)        \nonumber \\
& & + \pi^* \left( (\omega_P + i\omega_Q)\wedge t^{0,1} \right) + \bp\pi^*u^{1,1}     \nonumber \\
& \cong & \bar{\rho}\wedge\pi^*t^{1,1} + c\rho\wedge\pi^*\bar{\Omega}^{0,2}_{K3} + \pi^* \left( (\omega_P + i\omega_Q)\wedge t^{0,1} \right) ,
\eea
where $c$ is constant, $\bp t^{1,1} = 0$, and `$\cong$' means equal up to $\bp$-exact terms (which also identifies $\xi$ under $t^{1,1} \rightarrow t^{1,1} + \bp u^{1,0}$).  Notice that this last term is necessarily $\bp$-closed, but since $h^{1,2}(K3) = 0$, it is also $\bp$-exact and thus we have
\be
\xi \cong \bar{\rho}\wedge\pi^*t^{1,1} + c\rho\wedge\pi^*\bar{\Omega}^{0,2}_{K3} ,
\ee
which implies $h^{1,2}(K) = h^{1,1}(K3)+h^{0,2}(K3) = 21$.

Finally, we know from \cite{Strominger:1986uh} that $h^{3,0}(K) = 1$ and we can use this to fill out the Hodge diamond:

\be
\begin{array}{ccccccc}
 & & & 1 & & &   \\
 & & 0 & & 1 & &  \\
 & 1 & & 20 & & 1 &   \\
1 & & 21 & & 21 & & 1   \\
 & 1 & & 20 & & 1 &   \\
 & & 1 & & 0 & &    \\
 & & & 1 & & &
\end{array}
\ee

We do not expect that for a non-K\"ahler manifold the Hodge numbers will add up to the Betti numbers, and indeed they do not.  The Betti numbers were computed in \cite{Goldstein:2002pg} using the Gysin sequence.  For the sake of comparison they are
\be
\begin{array}{c|c|c}
 & \omega_Q \neq n\omega_P & \omega_Q = n\omega_P    \\
\hline b_0(K) & 1 & 1   \\
b_1(K) & 0 & 1     \\
b_2(K) & 20 & 21     \\
b_3(K) & 42 & 42  
\end{array}
\ee
for any $n\in\bb{Z}$.  We hope that this cohomology calculation will be useful when an index is found to count the number of moduli fields, but we leave this for future work.


\section{Discussion}
\label{discussion}

In \cite{Candelas:1985en} and \cite{Strominger:1986uh}, the supersymmetry constraints were satisfied in part by assuming no fermionic condensates.  In order to get the first image of the massless spectrum from compactification on the FSY geometry, we have counted the solutions of the variations of the gaugino that satisfy the linearized equations of motion from heterotic supergravity.  We found they are given by the cohomology of forms valued in a vector bundle using the gauge connection.  The ability to choose a trivial bundle relies on $c_2(TK) = 0$, which is true for the GP manifold.  Taking the gauge bundle to be trivial allowed us to relate the twisted Dolbeault cohomology to the ordinary Dolbeault cohomology of the GP manifold, which we then computed. 

This counting is far from a full treatment of the effective action or even the massless spectrum resulting from compactification on the FSY geometry.  First of all, one must include the new light modes arising from toroidal compactifications.  After including these new fields, one way to get an upper bound on the number of massless fields would be to count the solutions of the variations of all the fermions that satisfy the linearized equations of motion.  Unfortunately, these are complicated, coupled differential equations and perhaps no simpler to solve than other potential methods for addressing aspects of the effective action, such as trying to understand the moduli space of FSY geometries.\footnote{For example, in \cite{Becker:2005nb} the authors explored variations of the supersymmetry constraints under a set of simplifying assumptions.}  The reason we expect this to give only an upper bound is because we have ignored quartic fermionic terms in the action, higher order $\alpha'$ corrections, and a conjectured superpotential (see \cite{Becker:2003dz}, for example), all of which we expect to impose additional constraints and decrease the number of true massless fields.

There are many exciting topics yet to be explored in the realm of torsional compactifications of the heterotic string, the four-dimensional effective action being one of the ultimate goals.  The moduli space would also be interesting since we expect that the inclusion of flux in the heterotic theory will lift most of the moduli that we have in Calabi-Yau compactifications.  Understanding the moduli space could then help in understanding the four-dimensional effective action.  It would also be interesting to be able to compare the effective action for these heterotic compactifications to the type IIB dual that was studied in \cite{Frey:2003sd}.  Since the FSY geometry is the first smooth construction satisfying the $\mc{N}=1$ supersymmetry constraints derived from the supergravity approximation,\footnote{An orbifold limit of a torsional $T^2$ bundle over $K3$ was considered in \cite{Dasgupta:1999ss} by duality chasing.} the time is ripe for studying flux compactifications of the heterotic string; we hope the reader has gained some interest in studying these partially-forgotten topics!

\section*{Acknowledgements}

We would like to thank A. Strominger and A. Adams for their deep insights and unrelenting patience.  We would also like to thank M. Becker, M. Ernebjerg, S. Prokushkin, A. Simons, L.-S. Tseng, and S.-T. Yau, for many enlightening conversations.  This work was supported in part by DOE grant DE-FG02-91ER40654.


\begin{appendix}

\section{Einstein/String Frame Actions and EOM's}
\label{string-frame}

We start with the ten-dimensional action in Einstein-frame, as is found in Chapter 13 (p325) of \cite{Green:1987mn}, with the substitution $\phi_{GSW} = \exp[\phi/2 + 2 \ln(\kappa/g)]$:
\bea
L^{(E)}_{Het} & = & -\textstyle\frac{1}{2}\sqrt{-G}\Big[ \frac{1}{\kappa^2} R + \frac{1}{2\kappa^2}D_M \phi D^M \phi 
+ \frac{1}{2\kappa^2}e^{-\phi/2}\Tr(F^2) 
+ \frac{3}{4\kappa^2}e^{-\phi}H^2      \nonumber  \\
& + & \textstyle\bar{\psi}_M \Gamma^{MNP}D_N \psi_P 
+ \bar{\lambda} \Gamma^M D_M \lambda 
+ \Tr(\bar{\chi}\Gamma^M D_M \chi)      \nonumber \\ 
& + & \frac{1}{\sqrt{2}}\bar{\psi}_M \Gamma^N \Gamma^M \lambda D_N \phi
- \textstyle \frac{1}{8} e^{-\phi/2} \Tr(\bar{\chi}\Gamma^{MNP}\chi)H_{MNP}         \nonumber \\ 
& + & \frac{1}{2} e^{-\phi/4} \Tr(\bar{\chi}\Gamma^M\Gamma^{NP}(\psi_M + \frac{\sqrt{2}}{12}\Gamma_M \lambda)F_{NP})       \nonumber \\
& - & {\textstyle \frac{1}{8}e^{-\phi/2}}\big( \bar{\psi}_M\Gamma^{MNPQR}\psi_R + 6\bar{\psi}^N\Gamma^P\psi^Q - \sqrt{2}\bar{\psi}_M\Gamma^{NPQ}\Gamma^M\lambda\big) H_{NPQ}     \nonumber \\
& + & (\textrm{Fermions})^4 \Big] ,
\eea
where $D_M$ is the Levi-Civita connection plus the gauge connection.  We know that if we rescale the metric $G_{MN} \rightarrow e^{p\phi}G_{MN}$, then $\sqrt{-G}\rightarrow e^{pd\phi/2}\sqrt{-G}$ and
\be
R \rightarrow e^{-p\phi}\left( R - p (d-1) D^2 \phi - p^2\frac{(d-1)(d-2)}{4}D_M \phi D^M \phi \right)
\ee
for GSW conventions.

Since we want an overall factor of $e^{-2\phi}$ in front in string frame, we must choose $p=-\frac{1}{2}$.  Under this rescaling, we have
\be
\begin{array}{cc} G^{(E)}_{MN} = e^{-\phi/2}G_{MN}, \quad  & R^{(E)} = e^{\phi/2} (R - {\textstyle\frac{9}{2}} D_M\phi D^M \phi ),   \\
\lambda^{(E)} = e^{\phi/8} \lambda ,     &    \chi^{(E)} = e^{\phi/8} \chi ,   \\
\psi^{(E)}_M = e^{-\phi/8} \psi_M ,     &     \Gamma_M^{(E)} = e^{-\phi/4}\Gamma_M ,      \\
\epsilon^{(E)} = e^{-\phi/8}\epsilon ,     & 
\end{array}
\ee
where $\epsilon$ is the spinor appearing in the supersymmetry variations.  The Levi-Civita connection has additional terms depending on derivatives of $\phi$:
\be
\begin{array}{c}
D^{(E)}_M \lambda = D_M \lambda + {\textstyle\frac{1}{8}}\Gamma^N_{\phantom{N}M} \lambda D_N\phi ,     \\
D^{(E)}_M V_N = D_M V_N + {\textstyle\frac{1}{4}}\big( V_M D_N \phi + V_N D_M \phi - G_{MN} V^P D_P\phi \big) ,   \\
\Gamma^{(E)M}_{\phantom{(E)}NP} = \Gamma^M_{NP} - {\textstyle\frac{1}{4}}\big( \delta^M_N D_P\phi + \delta^M_P D_N\phi - G_{NP} D^M\phi \big) ,
\end{array}
\ee
where $V_M$ is a spacetime 1-form.

After some simplification, the string-frame action is
\bea
L^{(S)}_{Het} & = & -\textstyle \frac{1}{2} e^{-2\phi}\sqrt{-G} \Big[  \frac{1}{\kappa^2} R 
- \frac{4}{\kappa^2}D_M\phi D^M\phi 
+ \frac{1}{2\kappa^2} \Tr(F^2) 
+ \frac{3}{4\kappa^2} H^2      \nonumber  \\
& + & \textstyle \bar{\psi}_M \Gamma^{MNP} D_N \psi_P 
+ \bar{\psi}_M \Gamma^{MP}\Gamma^N \psi_P D_N\phi 
+ \bar{\lambda}\Gamma^M D_M \lambda     \nonumber \\
& - & \bar{\lambda}\Gamma^M\lambda D_M\phi    
+ \textstyle \Tr\big( \bar{\chi}\Gamma^M D_M\chi - \bar{\chi}\Gamma^M \chi D_M\phi \big)       \nonumber \\ 
& + & \frac{1}{2}\Tr\big( \bar{\chi}\Gamma^M\Gamma^{NP} (\psi_M + \frac{\sqrt{2}}{12}\Gamma_M\lambda) F_{NP} \big)     
+ \textstyle \frac{1}{\sqrt{2}}\bar{\psi}_M\Gamma^N\Gamma^M\lambda D_N\phi     \nonumber \\
& - & \frac{1}{8}\Tr\big( \bar{\chi}\Gamma^{MNP}\chi \big) H_{MNP} 
- \frac{1}{8}\Big( \bar{\psi}_M\Gamma^{MNPQR}\psi_R     
+ \textstyle 6\bar{\psi}^N\Gamma^P\psi^Q      \nonumber \\
& - & \sqrt{2}\bar{\psi}_M\Gamma^{NPQ}\Gamma^M\lambda \Big) H_{NPQ}  +   (\textrm{Fermions})^4 .
\eea

\section{Useful Relations}

\subsection{SUSY Implications}
\label{susy-eqns}

A few things to note.  First, working in string-frame implies that
\be
D_\mu \eta = 0.
\ee
Second, for $D_M\Omega$ we have
\bea
D_\mu \Omega_{abc} & = & 0 ,       \nonumber \\
D_d \Omega_{abc} & = & -3(D_d\phi)\Omega_{abc} ,        \nonumber \\
D_{\bar{a}}\Omega_{abc} & = & -(D_{\bar{a}}\phi)\Omega_{abc} ,
\eea
all of which follow from the fermionic supersymmetry variations, as do
\bea
\label{eqn:H-phi-relations}
\frac{3}{2}H_{\bar{a}d}^{\phantom{ad}d} & = & D_{\bar{a}}\phi ,      \nonumber \\
\frac{3}{2}H_{ad}^{\phantom{ad}d} & = & - D_a\phi ,      \\
4\gamma^m D_m\phi\eta & = & H_{mnp}\gamma^{mnp}\eta .     \nonumber
\eea
See \cite{Strominger:1986uh} for details.

\subsection{A Note about Non-K\"ahler Manifolds}
\label{non-kahler}

One of the drawbacks to working with non-K\"ahler geometries is that, contrary to one's naive expectation,
\be
\nabla_m (\Omega_{abc}\gamma^{abc}) \neq \gamma^{abc}\nabla_m\Omega_{abc}.
\ee
This arises from the fact that we are only summing over holomorphic indices and not all real indices; thus, the complex structure is implicitly used and we recall that, unlike in the K\"ahler case, the complex structure is not covariantly constant with respect to the Levi-Civita connection.  To use the product rule, we must write everything in real indices.

Define
\be
P_{\pm m}^{\phantom{\pm m}n} \equiv \frac{1}{2}(\delta_m^{\phantom{m}n} \mp iJ_m^{\phantom{m}n})
\ee
so that $P_{+a}^{\ph{+a}b} = \delta_a^{\ph{a}b}$, $P_{-\bar{a}}^{\ph{-a}\bar{b}} = \delta_{\bar{a}}^{\ph{-a}\bar{b}}$, and all other components are zero.  Since $\nabla^{(H)}_m J = 0$, we find
\be
\nabla_m P_{\pm n}^{\ph{\pm n}p} = \mp\frac{3i}{4} \left( H^p_{\ph{p}ms} J_n^{\ph{n}s} - H^s_{\ph{s}mn}J_s^{\ph{s}p} \right).
\ee

Thus,
\bea
\nabla_m (C_{abc}\gamma^{abc}) & = & \nabla_m \left(C_{pst}\gamma^{nqr}P_{+n}^{\ph{+n}p}P_{+q}^{\ph{+q}s}P_{+r}^{\ph{+r}t} \right)       \nonumber \\
& = & \gamma^{abc}\nabla_m C_{abc} + 3C_{pbc}\gamma^{nbc}\nabla_m P_{+n}^{\ph{+n}p}     \nonumber \\
& = & \gamma^{abc}\nabla_m C_{abc} + \frac{9}{2}C_{abc}\gamma^{\bar{a}bc}H^a_{\ph{a}m\bar{a}}.
\eea
Similarly,
\be
\label{eqn:C2-relation}
\nabla_m (C_{ab}\gamma^{ab}) =  \gamma^{ab}\nabla_m C_{ab} + 3C_{ab}\gamma^{\bar{a}b}H^a_{\ph{a}m\bar{a}}
\ee
and
\be
\nabla_m (C_a\gamma^a) =  \gamma^a\nabla_m C_a + \frac{3}{2}C_a\gamma^{\bar{a}}H^a_{\ph{a}m\bar{a}}.
\ee

\section{Derivation of (\ref{eqn:simplified-gaugino})}
\label{derivation}

When we consider a gauge bundle with structure group $G$ and embed this into a larger group $H$ ($E_8\times E_8$ or $SO(32)$), the adjoint of $H$ decomposes into a sum containing the adjoint of $G$.  Since we focus on variations of the gaugino orthogonal to the adjoint of $G$, the linearized gaugino equation of motion becomes
\be
0 = 2\Gamma^M D_M \chi - 2 \Gamma^M\chi D_M\phi - \frac{1}{4}\Gamma^{MNP}\chi H_{MNP} ,
\ee
where we have suppressed gauge indices.

The most general Ansatz for the variation of the gaugino is
\be
\delta\chi = \epsilon_- \otimes \left( C\eta + C_{ab}\gamma^{ab}\eta \right) - \epsilon_+ \otimes \left(\bar{C}\eta^* + \bar{C}_{\bar{a}\bar{b}}\gamma^{\bar{a}\bar{b}}\eta^* \right),
\ee
where $C,C_{ab}\in \Omega^*(K;V)$ and $\epsilon_\pm$ are covariantly constant spinors on $\mathcal{M}_4$ with chiralities $\pm 1$ respectively.  To simplify the gaugino equation of motion, we note that 
\bea
\nabla_a \eta = \tfrac{3}{4}H_{ab\bar{a}}\gamma^{b\bar{a}}\eta + \tfrac{3}{8}H_{a\bar{a}\bar{b}}\gamma^{\bar{a}\bar{b}}\eta = -\tfrac{3}{4}H_{ab}^{\ph{ab}b}\eta = \tfrac{1}{2}(\partial_a\phi)\eta     \\
\nabla_{\bar{a}}\eta = \tfrac{3}{4}H_{\bar{a}\bar{b}a}\gamma^{\bar{b}a}\eta + \tfrac{3}{8}H_{\bar{a}ab}\gamma^{ab}\eta = -\tfrac{1}{2}(\partial_{\bar{a}}\phi)\eta + \tfrac{3}{8}H_{\bar{a}ab}\gamma^{ab}\eta ,
\eea
which follow from $\nabla^{(H)}_m\eta = 0$, $\gamma^{\bar{a}}\eta=0$, and (\ref{eqn:H-phi-relations}).

We will focus on the terms involving $\epsilon_-$ as the others are obtained by complex conjugation and multiplication by the ten-dimensional charge conjugation operator.  Using the relations above, (\ref{eqn:C2-relation}), and the fact that the product of more than three gamma matrices with all holomorphic or anti-holomorphic indices is zero (since we work on a complex $3$-fold), we have
{\setlength\arraycolsep{0pt}
\bea
\Gamma^M D_M\delta\chi \Big|_{\epsilon_-} \, & = & \, ( \gamma^\mu D_\mu\epsilon_- )\otimes (C\eta + C_{ab}\gamma^{ab}\eta) - \epsilon_-\otimes \gamma^m D_m(C\eta + C_{ab}\gamma^{ab}\eta)     \nonumber \\
= -\epsilon_- &\otimes& \left\{ \gamma^a \left[ \left(D_a(C+C_{bc}\gamma^{bc})\right)\eta + (C+C_{bc}\gamma^{bc})D_a\eta \right] \right.     \nonumber \\
& & \left. + \gamma^{\bar{a}} \left[ \left(D_{\bar{a}}(C+C_{bc}\gamma^{bc})\right)\eta + (C+C_{bc}\gamma^{bc})D_{\bar{a}}\eta \right] \right\}     \nonumber \\
= -\epsilon_-&\otimes& \bigg\{ \Big[ \left( (D_aC)\gamma^a + (D_aC_{bc})\gamma^{abc} + 3C_{bc}H_{\ph{b}a}^{b\ph{a}c}\gamma^a \right)\eta      \nonumber \\
& & + \left( \tfrac{1}{2}(\partial_a\phi)C\gamma^a + \tfrac{1}{2}(\partial_a\phi)C_{bc}\gamma^{abc}\right) \eta \Big] + \Big[ \left( (D_{\bar{a}}C_{bc})\gamma^{\bar{a}}\gamma^{bc} \right)\eta    \nonumber \\
& & + \left( \tfrac{3}{8}C H_{\bar{a}ab}\gamma^{\bar{a}}\gamma^{ab} - \tfrac{1}{2}(\partial_{\bar{a}}\phi)C_{bc}\gamma^{\bar{a}}\gamma^{bc} \right) \eta \Big] \bigg\}     \nonumber \\
= -\epsilon_-&\otimes& \left\{ \left[ D_aC + \tfrac{3}{2}(\partial_a\phi)C - 3C_{bc}H_a^{\ph{a}bc} + 4D^b C_{ba}    \nonumber- 2(\partial^b\phi)C_{ba} \right]\gamma^a\eta \right.    \nonumber \\
& & \left. + \left[ D_aC_{bc} + \tfrac{1}{2}(\partial_a\phi)C_{bc} \right]\gamma^{abc}\eta \right\} .
\eea}

Similarly,
{\setlength\arraycolsep{0pt}
\bea
-\Gamma^M \delta\chi\partial_M&\phi& \Big|_{\epsilon_-} = \epsilon_-\otimes\left\{ (\gamma^a\partial_a\phi + \gamma^{\bar{a}}\partial_{\bar{a}}\phi)(C\eta + C_{bc}\gamma^{bc}\eta) \right\}    \nonumber \\
= \epsilon_- &\otimes& \left\{ (\partial_a\phi)C\gamma^a\eta + (\partial_a\phi)C_{bc}\gamma^{abc}\eta + 4(\partial^b\phi)C_{ba}\gamma^a\eta \right\}
\eea}
and
{\setlength\arraycolsep{0pt}
\bea
-\tfrac{1}{8}\Gamma^{MNP} H_{MNP} &\delta&\chi \Big|_{\epsilon_-} = \tfrac{1}{8}\epsilon_- \otimes \gamma^{mnp}H_{mnp} (C\eta + C_{ab}\gamma^{ab}\eta)    \nonumber \\
= \tfrac{3}{8}\epsilon_- &\otimes& \Big\{ C H_{ab\bar{a}}\gamma^{ab\bar{a}}\eta + C_{ab}H_{\bar{a}cd}\gamma^{\bar{a}cd}\gamma^{ab}\eta + C_{ab}H_{c\bar{a}\bar{b}}\gamma^{c\bar{a}\bar{b}}\gamma^{ab}\eta \Big\}    \nonumber \\
= \tfrac{3}{8}\epsilon_- &\otimes& \Big\{ \tfrac{4}{3}(\partial_a\phi)C\gamma^a\eta - \tfrac{4}{3}(\partial_a\phi)C_{bc}\gamma^{abc}\eta - \tfrac{4}{3}(\partial_{\bar{a}}\phi)C_{bc}\gamma^{\bar{a}}\gamma^{bc}\eta     \nonumber \\
& & + C_{ab}H_{c\bar{a}\bar{b}}\gamma^c\gamma^{\bar{a}\bar{b}}\gamma^{ab}\eta    \Big\}     \nonumber \\
= \tfrac{1}{2}\epsilon_- &\otimes& \Big\{ (\partial_a\phi)C\gamma^a\eta - (\partial_a\phi)C_{bc}\gamma^{abc}\eta - 4(\partial^b\phi)C_{ba}\gamma^a\eta     \nonumber \\
& & - 6 C_{bc}H_a^{\ph{a}bc}\gamma^a\eta \Big\} .
\eea}

Combining these, the gaugino equation of motion reduces to
\be
0 = -2\left( D_aC + 4D^bC_{ba} - 4(\partial^b\phi)C_{ba} \right) \gamma^a\eta - 2\left( D_aC_{bc} \right) \gamma^{abc}\eta
\ee
as claimed.

\end{appendix}

\bibliographystyle{h-elsevier}

\bibliography{paper4refs}

\end{document}